\documentclass[twocolumn]{nature}
\usepackage{setspace}
\usepackage{amssymb}
\usepackage{url}

\usepackage{graphicx}

\title{Sodium content as a predictor of the advanced evolution of
  globular cluster stars}%85 chars
\author{Simon W. Campbell$^{1}$, Valentina D'Orazi$^{2,1}$, David
  Yong$^{3}$, Thomas N. Constantino$^{1}$, John C. Lattanzio$^{1}$, Richard
  J. Stancliffe$^{3,4}$, George C. Angelou$^{1}$, Elizabeth C. Wylie-de
  Boer$^{3}$, Frank Grundahl$^{5}$}
\begin{document}

\maketitle

%================================================================
\begin{affiliations}
 \item Monash Centre for Astrophysics, School of Mathematical Sciences,
   Monash University, Melbourne, VIC 3800, Australia.
 \item Department of Physics \& Astronomy, Macquarie University, Balaclava
   Rd, North Ryde, Sydney, NSW 2109, Australia.
 \item Research School of Astronomy and Astrophysics, Australian National
   University, Weston, ACT 2611, Australia 
 \item Argelander-Institut f\"{u}r Astronomie, Universit\"{a}t Bonn, Auf dem
   H\"{u}gel 71, 53121 Bonn, Germany.
 \item Stellar Astrophysics Centre, Department of Physics and Astronomy,
   Aarhus University, Ny Munkegade 120, DK-8000 Aarhus C, Denmark.
\end{affiliations}
%================================================================

%Articles have a summary, separate from the main text, of up to 150 words,
%which does not have references, and does not contain numbers,
%abbreviations, acronyms or measurements unless essential. It is aimed at
%readers outside the discipline. This summary contains a paragraph 
% 1: (2-3 sentences) of basic-level INTRO to the field; 
% 2: A brief account of the BACKGROUND AND RATIONALE of the work; 
% 3: a statement of the main CONCLUSIONS (introduced by the phrase 'Here we
%    show' or its equivalent); 
% 4: and finally, 2-3 sentences putting the main FINDINGS into GENERAL CONTEXT so it is clear
%    how the results described in the paper have moved the field forwards.

%**Letter Summary Paragraph** (part of main text but independent/bold text):
%They begin with a fully referenced paragraph, ideally of about 200 words,
%but certainly no more than 300 words, aimed at readers in other
%disciplines. This paragraph starts with a 2-3 sentence basic introduction
%to the field; followed by a one-sentence statement of the main conclusions
%starting 'Here we show' or equivalent phrase; and finally, 2-3 sentences
%putting the main findings into general context so it is clear how the
%results described in the paper have moved the field forwards.

%\textbf{NATURE SUMMARY PARAGRAPH (200 words max)}\\ 
\begin{abstract} 
The asymptotic giant branch (AGB) phase is the final stage of nuclear
burning for low mass stars.  Although Milky Way globular clusters are now
known to harbour (at least) two generations of
stars\cite{carretta09,gratton12} they still provide relatively homogeneous
samples of stars which are used to constrain stellar evolution theory\cite{
  iben69,buonanno85,renzini88}. It is expected, and predicted by stellar
models, that the majority of cluster stars with masses around the current
turn-off mass will evolve through the AGB
phase\cite{kippenhahn90,landsman96}. However this has never been confirmed
observationally.  Here we show that all of the second generation stars in
the globular cluster NGC 6752 -- 70~\% of the cluster population -- fail to
reach the AGB phase.  Through spectroscopic abundance measurements we found
that every AGB star in our sample has a low sodium abundance, with [Na/Fe]
$\lesssim$ 0.18 dex, indicating they are exclusively first generation
stars.  The extremely high AGB failure rate implies that many clusters
cannot be reliably used for star counts to test stellar evolution
timescales if the AGB population is involved. A thorough investigation is
required to reconcile stellar theory with this observed phenomenon.

%  Stellar evolution
%calculations suggest that enhanced mass-loss may help models match the
%observational data.
%suggest that in order
%to model this phenomenon it is likely that improved mass-loss physics is
%needed for models of blue horizontal branch stars with surface temperatures
%higher than $\sim$ 11,500 K.
\end{abstract}
%%%%%%%%%%%%%%%%%

%================================================================
%\section{MAIN TEXT: 1500 words max, 1700 if Table in SI}
%

%
%
%\section*{Stellar Sample and Results}

\begin{figure}
  \centering
  \includegraphics[width=0.8\columnwidth]{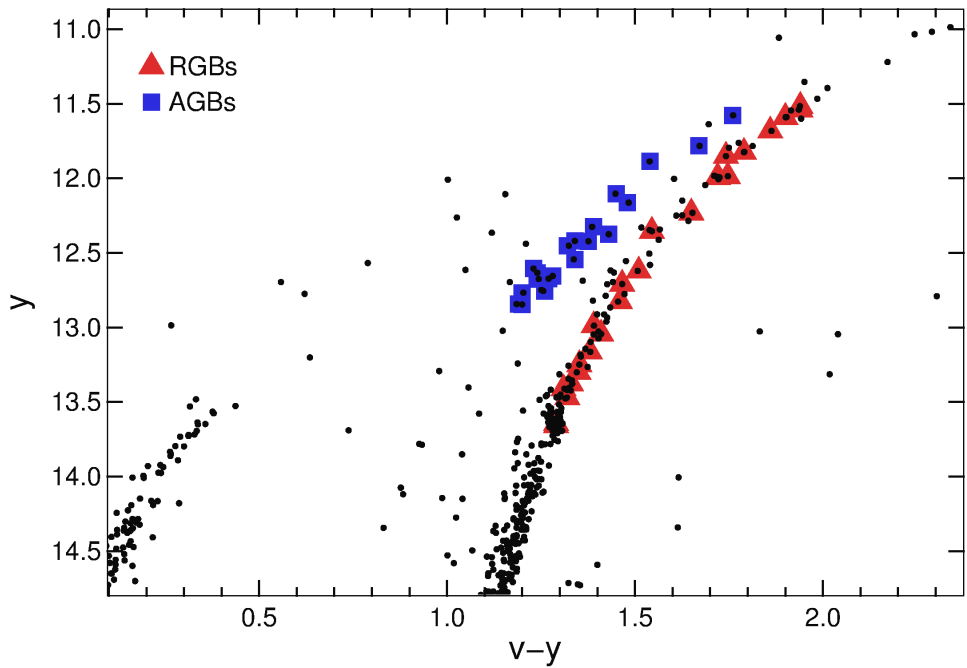}
  \caption{\textbf{Sample selection in the Str{\"o}mgren $uvby$
      colour-magnitude diagram of NGC 6752.} Small black dots show the
    whole photometric sample\cite{grundahl99}. Our AGB and red giant branch
    (RGB) stellar samples are shown as blue squares and red triangles
    respectively. Part of the horizontal branch can be seen at bottom left,
    at $y$ magnitudes $\gtrsim 13.5$.}
  \label{fig:sample}  
\end{figure}

We obtained high resolution spectra (R $\sim 24,000$) for a sample of 20
AGB stars and 24 red giant branch stars in the Galactic globular cluster
(GC) NGC 6752. The spectral coverage included the strong Na I doublet at
5680 \AA. In Figure \ref{fig:sample} we show the stellar sample. We
include red giant branch stars as a control group, since it has previously
been shown that this evolutionary population harbours the standard
abundance distributions, including the well-known Na-O anticorrelation
present in all GCs\cite{carretta07}.
%\textbf{red giant branch stars also have similar surface
%temperatures and gravity to AGB stars, making comparisons more direct.}
%------------------------
%------------------------
%
%------------------------
\begin{figure}
  \centering
  \includegraphics[width=0.7\columnwidth]{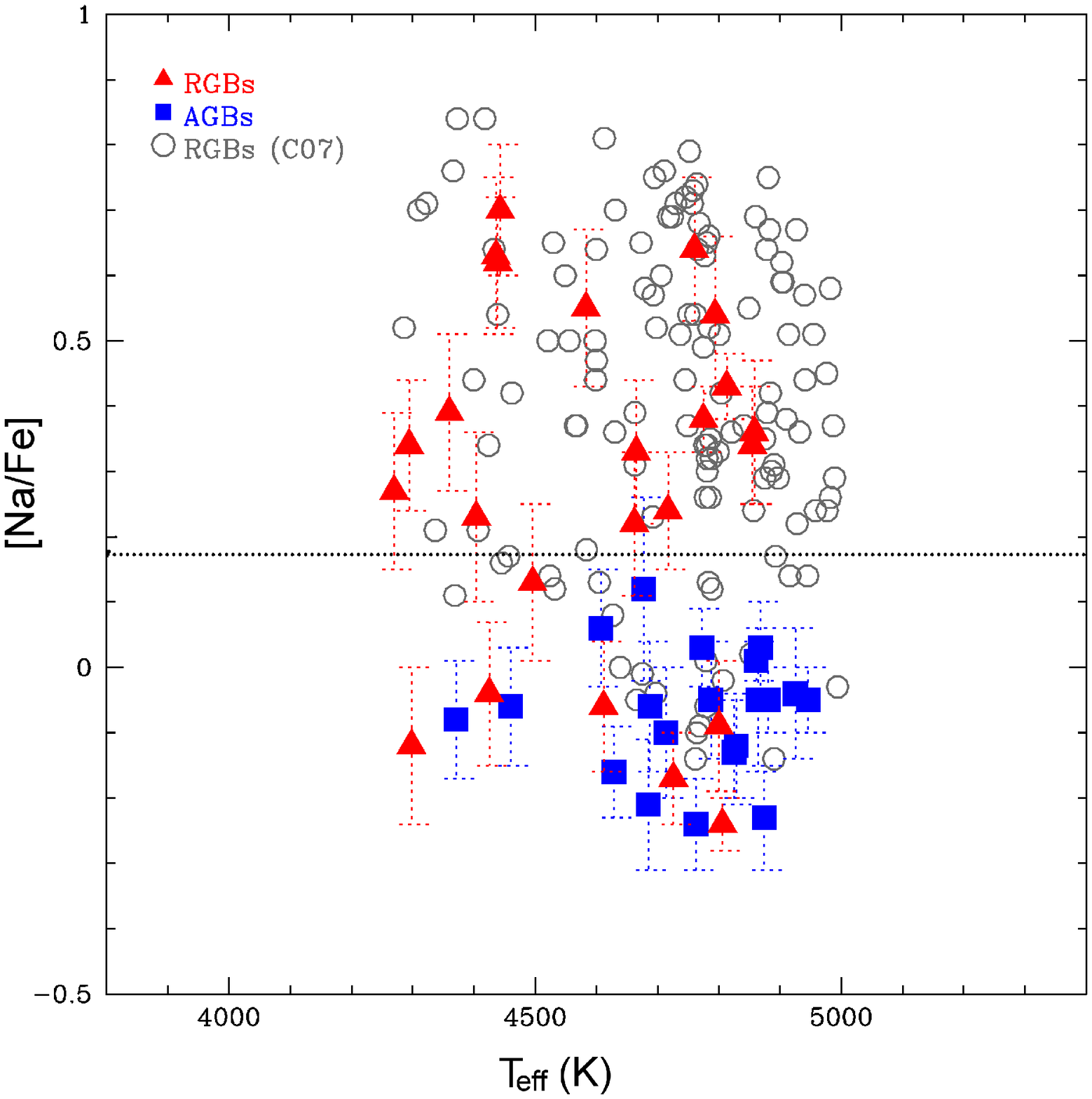}
  \caption{\textbf{Sodium abundance results for NGC 6752.} [Na/Fe] for our
    sample of red giant branch stars (red triangles, 24 stars) and AGB
    stars (blue squares, 20 stars) shown against stellar effective
    temperature $T_{eff}$ (also see Supplementary Table S1). For comparison
    the red giant branch results of a previous study (C07)\cite{carretta07}
    are included (grey open circles). The horizontal dotted line at [Na/Fe]
    $=0.18$ marks the upper envelope of AGB values, and divides Na-rich and
    Na-poor stars.  The spectroscopic observations were carried out with
    the Very Large Telescope under European Southern Observatory programme
    089.D-0038 (principal investigator SWC). The FLAMES/Giraffe HR11
    grating was used, with a spectral coverage $\lambda = 5597 \rightarrow
    5840$ \AA. Na abundances (assuming local thermodynamic equilibrium)
    were obtained from the strong Na I doublet at 5680 \AA, with the driver
    $abfind$ in MOOG\cite{sneden73} (2011 version) and the Kurucz model
    atmospheres with no overshooting\cite{kurucz93}.  Stellar parameters
    were derived in the following way: T$_{eff}$ values were calculated
    from a Str{\"o}mgren colour ($b$-$y$) calibration\cite{alonso01};
    gravities were then computed from stellar luminosities and the derived
    temperatures (with assumed stellar mass M $=0.8$ M$_\odot$ and distance
    modulus of $(m-M)_{V} = 13.30$), while microturbulence values $\xi$
    were obtained from the literature\cite{gratton96}. A metallicity of
    [Fe/H] $=-1.54$ dex\cite{carretta07} was adopted for all stars.
    Although the lines under scrutiny are known to be not largely affected
    by departures from local thermodynamic equilibrium (at most 0.15 dex),
    we applied non local thermodynamic equilibrium
    corrections\cite{gratton99} to our Na abundances.  Error bars show the
    random (internal) uncertainties (also see Supplementary Table S1). The
    uncertainties were estimated by adding, in quadrature, errors due to
    the equivalent width measurements and those related to stellar
    parameters. The latter were evaluated in the standard way, by varying
    one parameter at a time and inspecting the corresponding variation in
    the resulting abundances. We adopted errors of $\Delta$T$_{eff} = \pm
    30$ K, $\Delta\log$g$ = 0.1$, $\Delta\xi = 0.1$ km/s, and
    $\Delta$[Fe/H] $= \pm 0.05$ dex.}
  \label{fig:results}  
\end{figure}
%------------------------
%

Our sodium abundance results are shown in Figure \ref{fig:results}. The
red giant branch sample shows the usual spread in [Na/Fe] $=
\log_{10}(N_{Na}/N_{Fe})_{star} - \log_{10}(N_{Na}/N_{Fe})_{Sun}$ of
roughly 1 dex ($N_{x}$ is the number density of atoms of each elemental
species). On the other hand the AGB result is very striking -- every single
one of the AGB stars in our sample lies at the low end of the red giant
branch distribution. The upper envelope of the AGB sodium abundances is
located at about [Na/Fe] $=0.18$ dex. Interestingly this corresponds very
closely to a previous red giant branch study that defines the Na-poor,
first generation population as having [Na/Fe] $\lesssim 0.2$ dex (their
`Primordial' population)\cite{carretta09}. We find the proportions of
Na-poor to Na-rich red giant branch stars in our data to be 30:70.  This
also corresponds well to the roughly 30:70 proportions found
previously\cite{carretta09}. Thus, surprisingly, \emph{all} of our AGB
stars appear to be first generation stars, giving a first generation to
second generation ratio change from 30:70 in the red giant branch
population to 100:0 in the AGB population. The range in [Na/Fe] in our AGB
sample is very small, with a mean of $-0.07$ dex and a standard deviation
of 0.10 dex. This scatter is comparable to our internal uncertainties
(Fig. 2 and Supplementary Table S1), which suggests that the AGB stars may
have a uniform abundance of Na.
%Assuming the 30:70 Na-poor:Na-rich distribution as seen on the red giant branch, the probability
%of randomly selecting 20 Na-poor AGBs is extremely low, especially
%considering that we have derived Na abundances for almost all of the AGB
%stars in the region covered by the photometric dataset
%(Fig. 1\ref{fig:sample}).  

Our results indicate that the entire population of second generation stars,
having elevated levels of Na, must fail to enter the AGB phase.  This is a
very significant effect since the second generation population contains the
majority of the stars in NGC 6752 (70\%). Although two studies have
theorised that some AGB stars may not ascend the AGB in NGC
6752\cite{norris81}, including the study by our group\cite{campbell10},
this has been based on low-resolution cyanogen band strength
observations. It is known that these observations are affected by many
uncertainties\cite{yong08}, including the in situ variation of C and N
along the red giant branch\cite{smith92}.  Measurement of elemental Na is
more robust since Na is not affected by molecular band formation
uncertainties and stars of these masses ($\sim 0.8$ M$_\odot$) cannot alter
their Na abundances in situ. In particular a reduction in surface Na would
require very high temperatures at the base of the convective envelope. This
is only achieved in much more massive stars ($\gtrsim 4$ M$_\odot$), via
`Hot Bottom Burning' nucleosynthesis\cite{boothroyd93}. Thus the result
presented here is the first conclusive confirmation that stars of certain
chemical composition do not ascend the AGB. Moreover, we can readily
identify which stars avoid the AGB based on their Na content.

An obvious consequence of such a large proportion of stars avoiding the AGB
is that there should currently be many fewer stars in the AGB phase than
expected. A detailed study reporting star counts of GC populations finds a
value of $R_{2} = N_{AGB}/N_{HB} \sim 0.06$ for NGC
6752\cite{sandquist04}. This is one of the lowest $R_{2}$ values in their
GC sample. The GCs with the two highest $R_{2}$ values in their sample (M 5
and M 55) could be assumed to provide an upper limit to $R_{2}$ since
$R_{2}$ is fairly insensitive to metallicity, He abundance, and GC
age\cite{cassisi03}.  Interestingly this upper value is roughly 0.18,
i.e. a factor of 3 higher than that of NGC 6752. This is indeed consistent
with our result of $\gtrsim 70\%$ of stars not ascending the AGB. Current
model predictions for $R_{2}$ tend to be lower than 0.18, being around
$0.12 \rightarrow 0.15$\cite{cassisi01,cassisi03}, however the models are
known to suffer from significant uncertainties\cite{cassisi01}. We note
that the observed $R_{2}$ value for NGC 6752 is still at least a factor of
2 smaller than the model predictions.

%%%%%%%%%%%%%%%%%%%%%%%%%%%%%%%%%%%%%%%%%%%%%%%%%%%
%------------------------
\begin{figure}
  \centering
  \includegraphics[width=0.8\columnwidth]{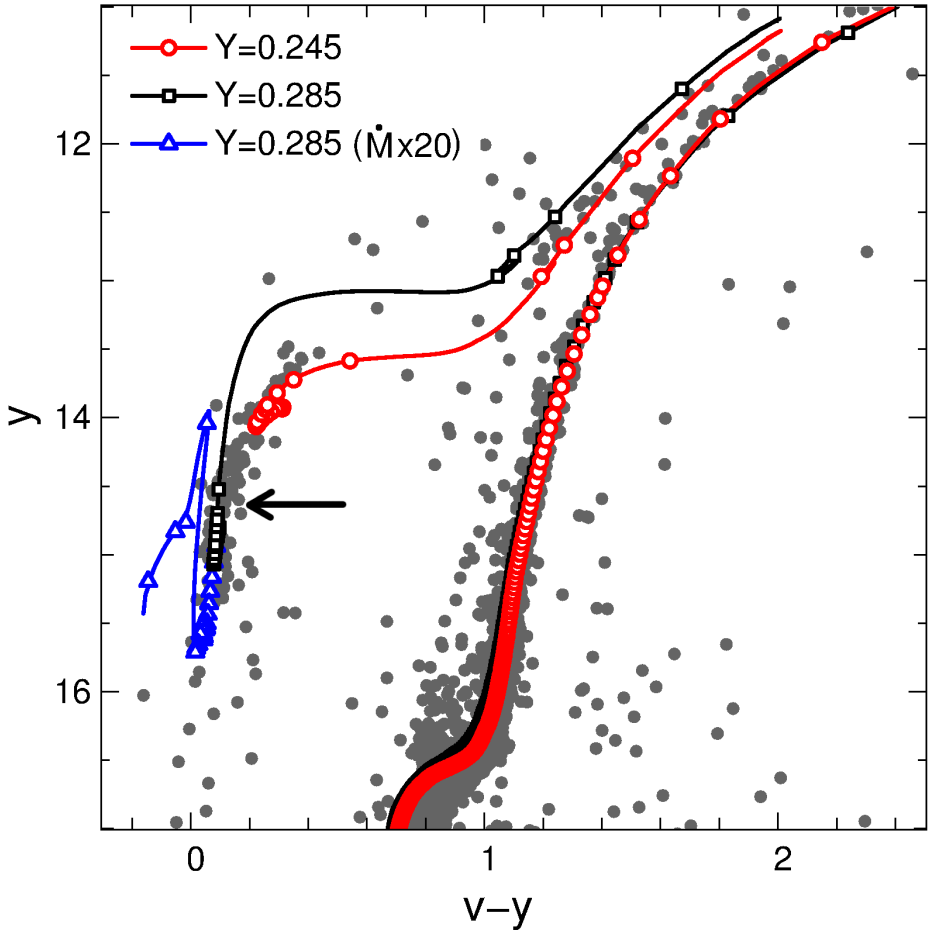}
  \caption{\textbf{Theoretical stellar model tracks overlain on the
      Str{\"o}mgren colour-magnitude diagram of NGC 6752.} The solid red
    line (with open circle symbols marking 5 Myr time intervals) is a model
    with an initial mass of 0.80 M$_\odot$ and a helium content of Y
    $=0.245$. This Y value matches that reported for the redder end of the
    horizontal branch\cite{villanova09}. This first generation model does
    indeed spend most of its horizontal branch evolution at the redder end
    of the horizontal branch. The solid black line (with open square
    symbols marking 5 Myr time intervals) is a model with an initial mass
    of 0.75 M$_\odot$ and an enhanced helium content of Y $=0.285$. This
    second generation model spends its horizontal branch evolution in a
    bluer part of the horizontal branch, but still ascends the AGB,
    contrary to the observational findings of the current study. The solid
    blue line (with open triangle symbols marking 5 Myr time intervals)
    shows the evolution of the $Y=0.285$ model with an ad-hoc 20-fold
    increase in mass loss rate ($\dot{M} = dM/dt$) initiated once the star
    settles on the horizontal branch. This model evolves downwards along
    the extreme blue end of the horizontal branch and fails to ascend the
    AGB. The arrow indicates the location of the Grundahl jump at $y =
    14.65$ (see text for details).  The stellar models were calculated
    using the Monash University stellar structure code
    MONSTAR\cite{campbell08}. The code has been recently updated with low
    temperature opacity tables which follow variations in C, N and
    O\cite{marigo09}. The usual Reimers mass loss rate\cite{reimers75} was
    used for the red giant branch and horizontal branch (with $\eta = 0.48$
    for the models with normal mass-loss). Transforms from theoretical
    luminosity-T$_{eff}$ plane to the colour-magnitude diagram have been
    made\cite{clem04}.}
    \label{fig:models}  
\end{figure}
%------------------------
%

Studies of the stage of evolution directly preceding the AGB, the
horizontal branch phase, can shed light on the AGB avoidance
phenomenon. One recent investigation into the Na abundances in a sample of
horizontal branch stars in NGC 6752 showed that the redder end of the
horizontal branch (NGC 6752 only has a blue horizontal branch) contains
only Na-poor stars\cite{villanova09} (see Supplementary Information section
S2 for more discussion). This implies that it is the bluer (presumably
Na-rich) horizontal branch stars that must avoid AGB ascent, leaving only
the redder, Na-poor horizontal branch stars to populate the AGB of NGC
6752. Combining this information with our ratio of Na-poor to Na-rich stars
(Figure 2) we can estimate the horizontal branch colour at which stars fail
to ascend the AGB. Interestingly we find an `ascension cut-off' colour that
coincides exactly with the colour for the Grundahl Jump. The Grundahl Jump
is a discontinuity in horizontal branch morphology which is seen in all GCs
studied to date whose horizontal branch extends beyond T$_{eff} \approx
11,500$ K\cite{grundahl99,momany02}.  Thus it appears that all stars bluer
than the Grundahl jump do not ascend the AGB, at least in NGC 6752. This
may represent further evidence that there is some fundamental change in the
stellar atmosphere structure and/or mass-loss physics occurring at the
Grundahl jump temperature\cite{grundahl99}.

We have calculated some stellar models to compare with the photometric
observations. The model results are shown in Figure \ref{fig:models}.  Our
first generation model populates the redder end of the horizontal branch,
before the Grundahl jump, as expected. It then continues to the AGB. The
He-rich second generation model (presumed to correspond to the Na-rich
population) populates the bluer end of the horizontal branch (after the
Grundahl jump), and also continues to the AGB. Thus our second generation
model cannot account for the lack of ascension of the Na-rich blue
horizontal branch stars in this part of the colour-magnitude diagram. We
note that an increased mass loss rate during the red giant branch phase
would result in a bluer zero-age horizontal branch star (see Supplementary
Information section S3 for more detail), so this can not be a solution
since the colour-magnitude diagram is clearly populated in this region. We
speculate that one solution to this problem may be that the horizontal
branch stars blueward of the Grundahl jump experience enhanced
mass loss. To test this we artificially increased the mass loss rate in the
second generation model during its horizontal branch phase by an ad-hoc
factor of 20 (Fig. 3\ref{fig:models}). Indeed this model can populate the
blue end of the horizontal branch and also fail to become an AGB star. We
note however that this test is purely hypothetical and, although this
result may provide a starting point, a thorough investigation into the
reasons behind the discordance between theory and observation is sorely
needed. There is currently no clear explanation for such a high proportion
of failed AGB stars.

%%%%%%%%%%%%%%%%%%%%%%%%%%%%%%%%%%%%%%%%%%%%%%%%%%

%\section*{Conclusion}
%
Finally, since GCs are often used to test stellar evolution theory, the
extremely high AGB failure rate reported here will affect any test which
uses star counts of AGB stars. This is true of the R-Method used to check
the lifetimes of various phases of evolution. In particular the $R'$, $R1$
and $R2$ values\cite{iben69,buonanno85,sandquist04} all involve the number
of AGB stars, so these values will be flawed (including the GC He values
inferred from them). This is particularly important if the GCs in question
have blue extensions to their horizontal branches, since it is the blue
horizontal branch stars that appear not to ascend the AGB. Star number
counts used to ascertain AGB lifetimes will also be misleading, unless the
proportion of AGB ascenders is known somehow, for example via an ascension
cut-off in Na abundance or horizontal branch colour.

\section*{References}
%=====================================================
%\bibliographystyle{apj}
%\bibliographystyle{ieeetr}
\bibliography{6752.bib}
%=====================================================

%**Acknowledgements should be brief, and should not include thanks to
%anonymous referees and editors, inessential words, or effusive comments. A
%person can be thanked for assistance, not “excellent” assistance, or for
%comments, not “insightful” comments, for example. Acknowledgements can
%contain grant and contribution numbers.

%**Author Contributions: authors are required to include a statement to
%specify the contributions of each co-author. The statement can be up to
%several sentences long, describing the tasks of individual authors referred
%to by their initials. See the authorship policy page for further
%explanation and examples.

%Author Information: Authors should include a set of statements at the end
%of the paper, in the following order:
%- Data deposition statement if appropriate, with the URL and relevant
%    numbers for public database accession.  
%- A sentence reading “Reprints
%    and permissions information is available at www.nature.com/reprints”.
%    Competing financial interests statement.  
%- A sentence reading
%    "Correspondence and requests for materials should be addressed to XX”,
%    where XX refers to one e-mail address.

%% Here is the endmatter stuff: Supplementary Info, etc.
%% Use \item's to separate, default label is "Acknowledgements"

\begin{addendum}
 \item[Supplementary Information] is linked to the online version of the
   paper at www.nature.com/nature.
 \item[Acknowledgements] We thank Dr. Yazan Momany of the European Southern
   Observatory (ESO, Chile) for providing his $UBV$ photometric dataset
   which is mentioned in Supplementary Information section
   S2. S.W.C. acknowledges support from the Australian Research Council's
   Discovery Projects funding scheme (project DP1095368). RJS is the
   recipient of a Sofja Kovalevskaja Award from the Alexander von Humboldt
   Foundation.  F.G. acknowledges funding for the Stellar Astrophysics
   Centre provided by The Danish National Research Foundation. The research
   is supported by the ASTERISK project funded by the European Research
   Council (Grant agreement no.: 267864). This work was based on
   observations made with ESO telescopes at the La Silla Paranal
   Observatory under programme ID 089.D-0038 and made extensive use of the
   SIMBAD, Vizier, 2MASS, and NASA ADS databases.
 \item[Author Contributions] S.W.C. designed and prepared the ESO Very Large
   Telescope (VLT) observing proposal, collected the spectroscopic data,
   and prepared the paper. V.D. reduced and analysed the spectroscopic
   data, and prepared the paper. D.Y. designed and prepared the ESO/VLT
   observing proposal and assisted in the paper
   preparation. T.N.C. calculated the stellar models and prepared figures
   for the paper. J.C.L. assisted in the preparation of the observing
   proposal and with the paper preparation. R.J.S., G.C.A. and
   E.C.W. assisted in the paper preparation and made preliminary
   observations with the Anglo-Australian Telescope. F.G. provided the
   $uvby$ photometric data for the AGB and red giant branch sample and
   assisted in the paper preparation.
 \item[Author Information] Reprints and permissions information is
   available at www.nature.com/reprints.  Correspondence and requests for
   materials should be addressed to simon.campbell@monash.edu.
\end{addendum}

%SUPPLEMENTARY INFORMATION
%-------------------------
%We recommend that, in all sections of the journal except Brief
%Communications Arising (for which SI is not permitted), SI in the following
%six 'flat' (text or figure) categories is combined into a single PDF, laid
%out as you wish readers, editors and peer-reviewers to download it. Each
%legend should be written directly beneath its figure.%

%When combining your 'flat' SI into a PDF for online publication, you should
%avoid inserting header or footer information (including pagination) into
%your documents, and you should leave sufficient space (approximately 2.5 cm
%at the top and bottom of the document) for standard Nature headers and
%footers to be inserted during in-house SI processing.%
%
%    1. Supplementary Figure(s) and Legend(s)
%    2. Supplementary Methods
%    3. Supplementary Table(s)
%    4. Supplementary Discussion
%    5. Supplementary Equation(s)
%    6. Supplementary Notes (including notes clarifying statistical analyses, acknowledgements, grant or other numbers)

%Some types of SI (listed below) are either best presented in editable
%format or cannot be presented as PDF for technical reasons. Please supply
%these types of SI in one of our allowable formats. They will be published
%with the PDF of the rest of your SI, downloadable as separate files.
%

%=====================================================
%=====================================================
\newpage
%\enlargethispage{8\baselineskip}
{\LARGE\noindent\textbf{{Supplementary Information}}}~
\begin{table}[!ht]
  \begin{flushleft}\textbf{Table S1: The stellar sample including atmospheric parameters and
    Na abundances.}\end{flushleft}\vspace{-10pt}
  \centering
  \singlespacing
  \scriptsize
    \begin{tabular}{cccccccc}
      \hline
        Type & ID    & T$_{eff}$ & $\log$ (g)  & $\xi$    & $\log$ (N$_{Na}$) &
        [Na/Fe]  & Error\\ \hline\hline
        AGB & FGJ000022 & 4607 & 1.414 & 1.765 & 4.85     & 0.06    & 0.09           \\ 
        AGB & FGJ000025 & 4371 & 1.146 & 1.851 & 4.71     & -0.08   & 0.09           \\ 
        AGB & FGJ000031 & 4460 & 1.285 & 1.806 & 4.73     & -0.06   & 0.09           \\ 
        AGB & FGJ000044 & 4629 & 1.537 & 1.725 & 4.63     & -0.16   & 0.07           \\ 
        AGB & FGJ000052 & 4787 & 1.740 & 1.660 & 4.74     & -0.05   & 0.09           \\ 
        AGB & FGJ000053 & 4688 & 1.636 & 1.693 & 4.73     & -0.06   & 0.10           \\ 
        AGB & FGJ000059 & 4772 & 1.719 & 1.666 & 4.82     & 0.03    & 0.06           \\ 
        AGB & FGJ000060 & 4685 & 1.654 & 1.687 & 4.58     & -0.21   & 0.10           \\ 
        AGB & FGJ000061 & 4714 & 1.689 & 1.676 & 4.69     & -0.10   & 0.10           \\ 
        AGB & FGJ000065 & 4677 & 1.540 & 1.724 & 4.91     & 0.12    & 0.14           \\ 
        AGB & FGJ000075 & 4763 & 1.764 & 1.652 & 4.55     & -0.24   & 0.07           \\ 
        AGB & FGJ000076 & 4881 & 1.850 & 1.624 & 4.74     & -0.05   & 0.05           \\ 
        AGB & FGJ000078 & 4868 & 1.855 & 1.623 & 4.82     & 0.03    & 0.07           \\ 
        AGB & FGJ000080 & 4829 & 1.843 & 1.627 & 4.67     & -0.12   & 0.08           \\ 
        AGB & FGJ000083 & 4825 & 1.849 & 1.625 & 4.66     & -0.13   & 0.08           \\ 
        AGB & FGJ000089 & 4861 & 1.868 & 1.618 & 4.80     & 0.01    & 0.04           \\ 
        AGB & FGJ000094 & 4925 & 1.937 & 1.596 & 4.75     & -0.04   & 0.10           \\ 
        AGB & FGJ000097 & 4946 & 1.978 & 1.583 & 4.74     & -0.05   & 0.05           \\ 
        AGB & FGJ000104 & 4874 & 1.907 & 1.606 & 4.56     & -0.23   & 0.08           \\ 
        AGB & FGJ201620 & 4864 & 1.938 & 1.596 & 4.74     & -0.05   & 0.11           \\ 
        RGB & FGJ000012 & 4270 & 1.062 & 1.878 & 5.06     & 0.27    & 0.12           \\ 
        RGB & FGJ000023 & 4360 & 1.181 & 1.840 & 5.18     & 0.39    & 0.12           \\ 
        RGB & FGJ000027 & 4425 & 1.290 & 1.805 & 4.75     & -0.04   & 0.11           \\ 
        RGB & FGJ000029 & 4298 & 1.102 & 1.865 & 4.67     & -0.12   & 0.12           \\ 
        RGB & FGJ000030 & 4294 & 1.070 & 1.876 & 5.13     & 0.34    & 0.10           \\ 
        RGB & FGJ000035 & 4439 & 1.353 & 1.784 & 5.41     & 0.62    & 0.10           \\ 
        RGB & FGJ000043 & 4443 & 1.359 & 1.782 & 5.49     & 0.70    & 0.10           \\ 
        RGB & FGJ000050 & 4404 & 1.267 & 1.812 & 5.02     & 0.23    & 0.13           \\ 
        RGB & FGJ000054 & 4496 & 1.487 & 1.741 & 4.92     & 0.13    & 0.12           \\ 
        RGB & FGJ000064 & 4436 & 1.353 & 1.784 & 5.42     & 0.63    & 0.12           \\ 
        RGB & FGJ000069 & 4583 & 1.587 & 1.709 & 5.34     & 0.55    & 0.12           \\ 
        RGB & FGJ000091 & 4665 & 1.776 & 1.648 & 5.12     & 0.33    & 0.11           \\ 
        RGB & FGJ000092 & 4612 & 1.711 & 1.669 & 4.73     & -0.06   & 0.10           \\ 
        RGB & FGJ000107 & 4662 & 1.822 & 1.633 & 5.01     & 0.22    & 0.11           \\ 
        RGB & FGJ000129 & 4717 & 1.939 & 1.596 & 5.03     & 0.24    & 0.09           \\ 
        RGB & FGJ000155 & 4726 & 1.992 & 1.579 & 4.62     & -0.17   & 0.07           \\ 
        RGB & FGJ000161 & 4775 & 2.052 & 1.559 & 5.17     & 0.38    & 0.05           \\ 
        RGB & FGJ000170 & 4794 & 2.083 & 1.549 & 5.33     & 0.54    & 0.12           \\ 
        RGB & FGJ000186 & 4800 & 2.117 & 1.538 & 4.70     & -0.09   & 0.10           \\ 
        RGB & FGJ000193 & 4806 & 2.134 & 1.533 & 4.55     & -0.24   & 0.04           \\ 
        RGB & FGJ000217 & 4813 & 2.161 & 1.524 & 5.22     & 0.43    & 0.05           \\ 
        RGB & FGJ000262 & 4855 & 2.252 & 1.495 & 5.13     & 0.34    & 0.09           \\ 
        RGB & FGJ000276 & 4858 & 2.260 & 1.492 & 5.15     & 0.36    & 0.11           \\ 
        RGB & FGJ200619 & 4760 & 1.940 & 1.595 & 5.43     & 0.64    & 0.11           \\
        \hline
    \end{tabular}
    \label{table}
\vspace{2mm} \\  \raggedright{\footnotesize The evolutionary status of each star is
  indicated in column 1. ID codes are designations of the current
  study. T$_{eff}$, $\log$ (g), and $\xi$ are the surface temperature,
  gravity, and microturbulence values used in the abundance determinations.
  $\log$ (N$_{Na}$) and [Na/Fe] are the final Na abundances. The final
  column shows the internal errors in [Na/Fe].}
\end{table}
%
%=====================================================
%
\newpage
\singlespacing
\noindent{\large\textbf{Discussion}}

%\noindent\textbf{S2. Relationship between horizontal branch morphology and
%  stellar composition}
\noindent\textbf{Relationship between horizontal branch morphology and
  stellar composition}
\vspace{-15pt}

\noindent It has long been speculated that the composition differences
between the first generation and second generation populations could have
an effect on the colour-magnitude diagram structure of
GCs$^{9}$. Recent work has begun to provide some evidence to
support this. For example, a new study on M4, which has both a red
horizontal branch and a blue horizontal branch, has shown that all red
horizontal branch stars in their sample are Na-poor, whilst all their blue
horizontal branch stars are Na-rich\cite{marino11}. They infer that the He
content must be different between the two Na populations since it is not
expected that Na (or N) could affect the position of stars in the
horizontal branch, while He can\cite{dantona02} (see Supplementary
Discussion S3 below).  As mentioned in the main text, a sample of horizontal
branch stars from the redder end of the horizontal branch of NGC 6752 (it only
has a blue horizontal branch) was shown to exclusively contain Na-poor
stars$^{17}$ (that study also reports one star with elevated Na
abundance, but, as noted by the authors, the star has evolved off the
horizontal branch and probably started from a much bluer position).  The
same stars have a uniform He abundance that is consistent with Big Bang
theory predictions (Y $=0.245$), as expected for a first generation
population. Thus it appears that the bluer (presumably Na-rich) horizontal
branch stars must avoid AGB ascent -- leaving only the redder, Na-poor
horizontal branch stars to populate the AGB of NGC 6752.
\vspace{-18pt}

As also mentioned in the main text, if we combine this information with our
estimate of the proportion of stars that do not ascend the AGB, we can
estimate what horizontal branch colour delineates the border between the
two groups. Since our $uvby$ colour-magnitude diagram dataset is not
complete at the bluest end of the horizontal branch, we obtained a very
high quality $UBV$ photometric dataset$^{19}$ for this purpose,
courtesy of Dr. Yazan Momany at the European Southern Observatory
(Chile). We counted horizontal branch stars in the $U$, $U-V$ plane,
starting at the red edge of the horizontal branch at $U-V = 0.25$. The
total number of horizontal branch stars was found to be 320. Thus we expect
the reddest 96 stars (30\%) to eventually ascend the AGB. We find that this
number of stars corresponds to an `ascension cut-off' in $U-V$ of
$-0.30$. Interestingly this is exactly the colour for the Grundahl jump, a
discontinuity in horizontal branch morphology which is seen in all GCs
studied to date whose horizontal branch extends beyond T$_{eff} \approx
11,500$ K$^{18,19}$.  Explanations of this discontinuity
include radiative levitation of elements heavier than carbon and nitrogen
in the high-temperature atmospheres of these stars$^{18}$, or the
combination of post-zero-age horizontal branch evolution and diffusion
effects$^{19}$.  At face value, it appears that all stars bluer
than the Grundahl jump do not ascend the AGB, at least in NGC 6752. This
may represent further evidence that there is some fundamental change in the
stellar atmosphere structure and/or mass-loss physics occurring at the
Grundahl jump temperature$^{18}$.  We note that it is these
extremely blue horizontal branch stars that are considered to be the source
of excess UV flux in the spectra of elliptical galaxies\cite{greggio90} as
well as Galactic and extra-galactic GCs\cite{dalessandro12}.

%\noindent\textbf{S3. Stellar model experiments}
\vspace{-8pt}
\noindent\textbf{Stellar model experiments}
\vspace{-15pt}

\noindent
Theoretical models show that for a given metallicity and core mass at the
tip of the red giant branch the position of a star along the horizontal
branch is determined by the mass of the hydrogen-rich envelope. The lower
the envelope mass, the bluer the star will be. If the envelope mass is
extremely low ($\lesssim 10^{-2}$ M$_\odot$)\cite{dcruz96}, a horizontal
branch star will not ascend the AGB. These stars instead evolve directly to
the WD cooling track. A horizontal branch star can have such a low envelope
mass if it suffered extra mass-loss during the preceding red giant branch
phase, or if it formed with an elevated helium abundance. In the latter
case the higher He affects the evolution of the star such that it arrives
on the horizontal branch with a lower total mass (for a given GC age). In
the former case the mechanisms that might affect the mass loss rates are
unknown, although rotation is a possibility. In both cases the stars
populate the blue end of the horizontal branch\cite{dantona02}.
\vspace{-18pt}

Our stellar model experiments (see Figure 3 in main text) are based on
standard assumptions for mass-loss rates, cluster age, and a possible
helium variation between the two generations of $\delta$Y $=0.04$ (although
we note a very recent study has just reported stronger constraints on
Y)\cite{milone13}. Different assumptions of mass-loss rate or cluster age
could produce bluer (or redder) horizontal branch morphologies. Our
simulations show that standard models cannot reproduce the observations and
thus non-standard/improved models are needed. We are currently working on
improved horizontal branch models and expect that other groups will also
investigate the very high AGB failure rate phenomenon reported in this
Letter. Finally we note that our rough test of enhanced mass-loss on the
horizontal branch (Figure 3 in main text) resulted in a quite different
evolution in the CMD -- the star with a 20-fold increase in mass-loss
evolved `downwards' (towards lower luminosities), practically along the
zero-age horizontal branch line, then spent a significant amount of time at
higher luminosities after leaving the zero-age line. This different
evolution should be taken into account in future theoretical and
observational investigations. We thank a thoughtful referee for inspiring
this final paragraph of discussion.

%=====================================================
%%
%% TABLES
%%
%% If there are any tables, put them here.
%%
%=====================================================

%\input{supp.tex}
%\includepdf[pages=-]{SI-3.pdf}
%\includepdf[pages=-]{figs.pdf}

%\begin{figure*}[t]
%\centering
%\includegraphics[width=1.0\textwidth]{figs.pdf}
%\end{figure*}

\end{document}